# Cardiac segmentation for ventricular tachycardia radioablation and radiotherapy patients with cardiac implants.

**Authors:**

Nicholas Summerfield, PhD[1,2]

Dustin Jacqmin, PhD[1]

Chase Ruff, M.S[1,2]

Andrew M. Baschnagel, MD[1]

Adam Burr, MD, PhD[1]

Michael Bassetti, MD, PhD[1]

Ryan Kipp, MD[3]

Matthew Kalscheur, MD[3]

Patrick M. Hill, PhD[1]

Ming Dong, PhD[4]

Carri Glide-Hurst, PhD[1,2]

**Affiliations:**

1. Department of Radiation Medicine, University of Wisconsin-Madison, Madison, Wisconsin
2. Department of Medical Physics, University of Wisconsin-Madison, Madison, Wisconsin
3. Division of Cardiovascular Medicine, Department of Medicine, University of Wisconsin–Madison, Madison, WI
4. Department of Computer Science, Wayne State University, Detroit, Michigan

**Corresponding Author:** Carri Glide-Hurst, glidehurst@humonc.wisc.edu, Dept of Radiation Medicine, 600 Highland Avenue, K4/B100-0600, Madison, WI 53792

**Statistical Author:** Nicholas Summerfield, nsummerfield@wisc.edu, Dept of Radiation Medicine, 600 Highland Avenue, K4/B100-0600, Madison, WI 53792

**Funding statement:** Work reported in this publication was supported in part by the National Cancer Institute of the National Institutes of Health under award numbers R01HL153720 (PI: Carri Glide-Hurst) and T32CA009206. The content is solely the responsibility of the authors and does not necessarily represent the official views of the National Institutes of Health.

**Data sharing statement:** Data is not available at this time.

**Acknowledgements:** None.

**Disclosures:** C.G.-H. reports research collaborations with Modus Medical, Inc, RaySearch, Leo Cancer Care, and Medscint, outside of the submitted work. D.J. reports consulting fees from ASTRO and holds an unpaid leadership role in a specialty meeting of AAPM, outside the submitted work. A.Baschnagel reports research collaboration or funding with the Wisconsin Partnership Program, Partnership Education and Research Committee, and Wisconsin Alumni Research Foundation, outside of the submitted work. A.Burr reports research collaboration, funding, or consultation with GE Healthcare and Siemens outside of the submitted work. R.K. reports consulting fees from BioSense Webster, participation on board committees for Murj, Inc. and Varian, Inc., and other financial or non-financial interests in BioSense Webster, Medtronic, and Murj, Inc., outside of the submitted work. M.K. reports grants or contracts from GE Healthcare, outside of the submitted work. P.H. reports patent royalties and leadership for the American College of Radiology, outside of the submitted work. N.S., C.R., M.B., and M.D. have no disclosures.

## Abstract

**Purpose:** Deep-learning (DL) auto-segmentation enables cardiac substructure (CS) delineation for advanced cardiac-sparing radiation therapy (RT). Patients presenting for RT may have implantable devices such as cardioverter-defibrillators (ICDs), which induce metal artifacts. Furthermore, stereotactic body RT (SBRT) is used to treat ventricular tachycardia (VT), a population frequently presenting with ICDs. We apply a DL model for RT patients with ICDs or undergoing VT-SBRT, enabling auto-segmentation and substructure-level dose analysis.

**Methods:** Nineteen patients with ICDs who underwent CT simulation (CT-SIM) and received VT-SBRT were evaluated. A previously validated multi-modal nnU-Net-based segmentation method was retrained using a highly-curated CT-SIM cohort, supplemented with ICD images for training (ICD-specific, 8; total, 33) and validation (ICD-specific, 1; total, 6), with a hold-out ICD test set (n=10), to predict 20 CS including whole heart (WH), chambers, great vessels, coronary arteries, valves, and conduction nodes. Additional CT-SIM datasets included cancer-RT patients with (n=9) and without (n=10) ICDs evaluate generalizability. The retrained ICD-supplemented model was compared against a non-ICD model using Dice similarity coefficient (DSC), 95% Hausdorff distance (HD95), and Wilcoxon signed-rank test ($p<0.05$). Dosimetric evaluation (Dmean, maximum (D0.03cc)) was performed for VT-SBRT CS.

**Results:** For VT-SBRT, the ICD-supplemented model achieved average DSC/HD95 of 0.71±0.19/10.7±22.5mm respectively, outperforming ($p<0.05$) the non-ICD model. On CT-SIM images with ICDs, the model achieved an average DSC/HD95 of 0.70±0.22/7.9±5.1mm respectively, outperforming ($p<0.05$) the non-ICD model. On CT-SIM images without ICDs, the model achieved an average DSC of 0.74±0.17 (Non-ICD model, DSC, 0.75±0.16, $p<0.05$) and average HD95 of 5.5±3.1mm (Non-ICD model, HD95, 5.3±2.8mm). For VT-SBRT, CS dose depended on target location, with higher relative dose delivered to the WH, left/right ventricles, tricuspid valve, atrioventricular node, and left anterior descending artery than other substructures.

**Conclusion:** ICD-supplemented training improved CS segmentation despite metal artifacts without meaningfully degrading non-ICD performance, supporting auto-segmentation and substructure-level dosimetry in ICD populations.

## Introduction

Radiation included heart disease is a recognized complication of thoracic radiation therapy (RT)[1]. Yet, cardiac radiosensitivity is not uniform across the heart and conventional dose descriptors such as mean heart dose do not capture the heterogeneity of dose delivered to individual cardiac substructures (CS)[2]. Growing evidence suggests exposure of specific regions, including coronary arteries (CAs), chambers, heart base, and the conduction system may be more closely associated with post-RT toxicities such as ischemic events, induced arrhythmias, and heart failure than implementing whole-heart metrics alone[1,2]. Recent thoracic radiotherapy studies support more routine inclusion of CS in treatment planning and toxicity modeling, while emerging treatment planning studies suggest CS-sparing approaches are technically feasible with modern RT practices[3,4]. Accordingly, successful CS-sparing may help reduce the risk for clinically relevant late-effects, including induced coronary ischemic disease, arrythmias, and heart failure, while preserving baseline cardiac function[5].

The prevalence of patients presenting to the clinic with cardiac implantable electronic devices (CIEDs), such as implantable cardioverter-defibrillators (ICDs) or pacemakers, are becoming more common[6]. While these patients typically start with a weakened cardiac baseline reflective of their cardiovascular disease, CS-sparing techniques remain important to preserve remaining functionality and maximize potential post-RT outcomes. However, the presence of CIEDs or other surgical cardiac implants (i.e., valve replacement) perturbs cardiac anatomy with the presence of foreign, often high-density material that may introduce metal-induced artifacts in medical imaging. Furthermore, underlying structural heat diseases such as cardiomyopathy, heart failure, and myocardial infarctions may result in atypical cardiac anatomy that can introduce changes in wall thickness, dilation, scar formation, and ventricular shape distortion, that may result in deformed or displaced CS[7]. The presence of foreign material, imaging artifacts, and atypical cardiac anatomy challenges the ability for accurate image segmentation, particularly for CS to facilitate CS sparing techniques.

Another cardiac application implementing stereotactic body RT (SBRT) has introduced a noninvasive treatment option for patients with recurrent or refractory ventricular tachycardia (VT)[8,9] by delivering high, single fraction doses (25Gy) to portions of the left ventricle (LV). Early clinical evidence suggests that SBRT can reduce VT burden and ICD shocks as well as improve arrhythmia control in patients with limited treatment options[8]. Currently, SBRT treatment planning for VT does not routinely consider dose to the CS and patients typically always present with ICDs with high incidents of other surgical interventions and structural cardiac changes. A recent study by Miszczyk *et al.*[9] demonstrated initial feasibility of preferential sparing of the CAs over the planning target volume (PTV), successfully treating VT while sparing CAs. Yet, this practice is not routine and introduces additional contouring burden to the clinic. Incorporating CS into VT-SBRT for retrospective and prospective evaluations of precise CS dosimetry are needed to better understand treatment efficacy, toxicity, and dose-response relationships and explore opportunities for CS-sparing. This highlights the importance of minimizing the cardiac dose to preserve cardiac function. The same paper reports that cardiac function as evaluated by LVEF revealed no obvious deterioration after RT.

Deep learning (DL)-based CS auto-segmentation has shown strong performance on conventional radiotherapy datasets including simulation CT (CT-SIM) imaging[10]. However, these models are typically developed using patient populations without implantable devices, fewer imaging artifacts present, and healthier cardiac baselines (i.e., lack of cardiac structural changes)[10]. As a result, model performance may degrade when applied to these out-of-distribution yet clinically relevant cases[11]. Van der Pol *et al.*[12] recently demonstrated the feasibility of auto-segmentation in a large, multi-institutional VT cohort of patients with ICDs and leads by leveraging two models for CT-SIM datasets to target 16 CS. Yet, performance remained lower for small, challenging structures such as the cardiac valves and coronary arteries. As VT-SBRT becomes more broadly adopted and patient cohorts with CIEDs continue to grow, efficient DL methods that can accommodate these populations are needed.

To address this need, we developed a DL CS auto-segmentation model that can accommodate patients with ICDs considering two major use cases: VT-SBRT with substantial cardiac structural changes and external beam RT cohorts. The feasibility of demonstrating model use was applied to retrospectively assess dose to CS in patients treated with VT-SBRT to establish the technical groundwork for future substructure-specific dosimetric and clinical investigations.

## Methods

To fully train and evaluate the segmentation model, three different clinical datasets were implemented: a multi-modal, non-ICD dataset with patients who have no CIEDs and not diagnosed with structural heart disease, a CT-SIM dataset of VT patients undergoing SBRT with widespread ICDs and widespread structural heart disease, and a CT-SIM dataset of patients undergoing RT with cardiac-implants present including ICDs and other cardiac interventions such as artificial valves and septal defect repairs and some instances of structural heart disease. Images were manually labeled for 20 CS by an experienced cardiac annotator including whole heart (WH); chambers, left/right atrium/ventricles (LA, RA, LV, RV); great vessels (GVs), ascending aorta (AA), superior/inferior vena cava (SVC, IVC), pulmonary artery/veins (PA, PVs); coronary arteries (CAs), right (RCA), left main (LMCA), left anterior descending (LADA), and circumflex (LCx) arteries; valves, aortic (V-AV), pulmonic (V-PV), mitral (V-MV), tricuspid (V-TV) valves; and conduction nodes, sinoatrial (N-SA) and atrioventricular (N-AV) nodes following a consensus of published guidelines[13–16]. Expert consultation was performed for complex cases with a radiologist with cardiovascular subspecialty and 10+ years of experience.

## Non-ICD Dataset

A multi-modal dataset of 121 manually delineated, clinically-relevant images without ICDs, including CT-SIM, low-field magnetic resonance imaging coupled with linear accelerators (MR-Linac), and cardiac CT-angiography (CCTA) is retrospectively reviewed as previously described[17]. Within the CT-SIM arm, 40 images were split into training (n=25), validation (n=5), and testing (n=10) cohorts. The training and validation cohorts formed the baseline optimization dataset that provide additional fitting guidance and enable generalizability to CT-SIM images without ICDs used in the clinic. The test set was held independently to evaluate the final models on a cohort of CT images without ICDs as a reference performance baseline.

## VT-SBRT Dataset

Nineteen patients with ICDs who underwent SBRT (25 Gy, 1 fraction) for ischemic and non-ischemic recurrent VT at a single institution following prior interventions were retrospectively reviewed in this Institutional Review Board-approved study. CT-SIM imaging (120kVp, 100-200mAs, ~1mm in-plane resolution, 1-3mm slice thickness) was conducted on a Somatom Edge CT scanner (Siemens Healthineers, Malvern, PA) using respiratory gating (n=5) or 4D with compression (n=14), with (n=10) and without (n=9) intravenous contrast, and reconstructed with (n=9) and without (n=10) iterative metal artifact reduction (Siemens Healthineers, Malvern, PA). All patients underwent planning with electrophysiological and electroanatomic maps when available. PTV and scar locations were defined following published clinical trial guidelines[18]. The 19 patients are added to the existing Non-ICD CT-SIM cohort, resulting in an ICD-supplemented dataset. In total, eight patients were added to training (total, 33), one added to validation (total, 6), and ten were reserved as an ICD-specific testing dataset.

## Cardiac-Implant, Cancer-RT dataset

Ten patients with cardiac-implanted devices who underwent RT for primary or metastatic cancers were retrospectively reviewed. Treatment sites were widespread, but all patients underwent CT-SIM (120kVp, 82-1424mAs, ~1mm in-plane resolution, 1-3mm slice thickness) on either a Somatom Edge or X.Ceed CT scanner (Siemens Healthineers, Malvern, PA) that captured the full cardiac volume. Images were acquired using standard 3D acquisitions with compression (n=4), respiratory gating (n=4), or averaged 4D acquisitions (n=2), without intravenous contrast, and reconstructed with (n=5) and without (n=5) iterative metal artifact reduction. Nine patients had CEIDs including pacemakers, leadless pacemakers, and ICDs and one patient underwent treatment for ventral septal defect. All ten cases were reserved for model testing to determine applicability to a wider population of radiotherapy patients presenting with cardiac-implants or prior cardiac intervention that may affect the visualization of cardiac anatomy on CT-SIM.

## Deep Learning Methods

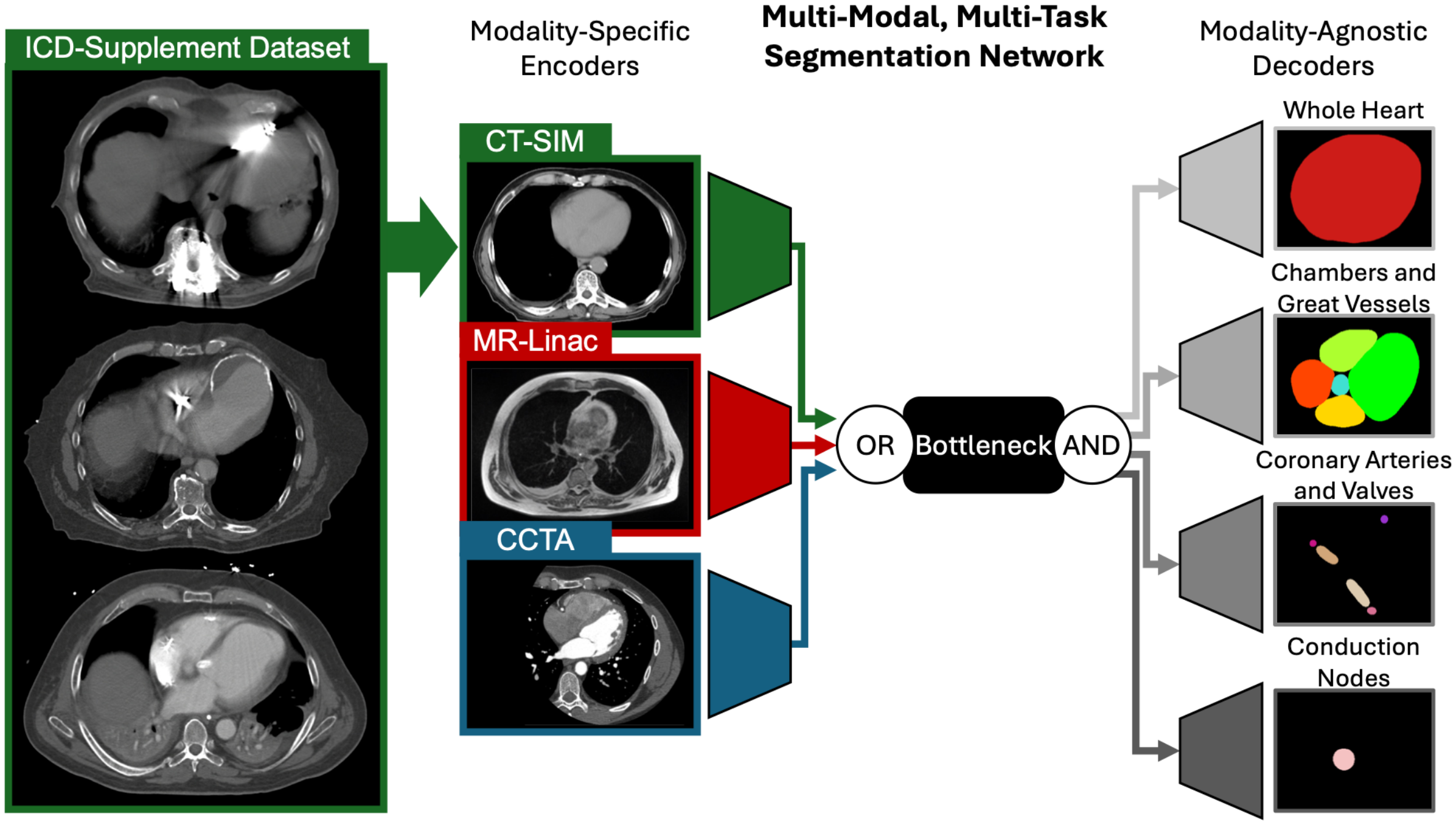

Figure 1: Implantable cardioverter-defibrillator (ICD) supplemented simulation CT (CT-SIM) cohort demonstrating metal artifacts, surgical intervention, and structural heart disease (left) and a previously described multi-modal, multi-task segmentation model for cardiac segmentation on CT-SIM, MR-Linac, and cardiac CT angiography (CCTA) images (right).

This work extends a previously developed and validated multi-modal, multi-task segmentation model capable of semantic segmentation for 20 CS on CT-SIM, MR-Linac, and CCTA images in a single framework[17] as shown in Figure 1. To improve model generalizability to this clinically complex population, the existing CT-SIM cohort from the multi-modal non-ICD dataset was supplemented with nine images from the VT-SBRT dataset. To assess the value of supplementing limited ICD-domain data with the broader curated CT-SIM cohort, three different models were compared:

(1) ICD-Supplemented model: the previously described[17] multi-modal, multi-task network trained with both the Non-ICD dataset and VT-SBRT dataset,

(2) ICD-Only model: the same network architecture as (1), configured for and trained on only VT-SBRT datasets, and

(3) Non-ICD model: the original model, as previously presented and validated[17], trained only on the Non-ICD dataset. This model served as a comparator to evaluate the added value of ICD-based training and to establish baseline performance on standard CT-SIM images with relatively typical cardiovascular anatomy.

The ICD-Supplemented and ICD-Only models were trained using the same preprocessing, model architecture, augmentation strategies, optimization framework, and compute environment as previously described[17]. Three different testing sets are evaluated: (1) VT-SBRT test set consisting of 10 VT-SBRT cases with ICDs, implant-related artifacts, variable contrast, and structural heart disease, used to evaluate model performance in an extreme setting, (2) Cardiac-Implant Cancer-RT test set including 10 CT-SIM cases for patients undergoing cancer-related RT with cardiac implants or prior cardiac surgical intervention used to assess generalizability to clinically relevant RT patients that present with implant-altered cardiac anatomy, and (3) a non-ICD test set that includes held-out CT-SIM cases without cardiac implants, used to assess maintained generalizability to most RT cases.

## Evaluation techniques

Model performance on each CT-SIM test set was measured quantitatively against the reference contours through the Dice Similarity Coefficient (DSC) and 95% Hausdorff Distance (HD95). Differences in CS segmentation performance between compared methods were assessed using two-tailed Wilcoxon signed-rank tests, with $p<0.05$ considered statistically significant. In cases where a structure was not predicted and HD95 was undefined, the value was imputed using the image diagonal following published guidelines[19]. Predicted and reference contours for the VT-SBRT and Cardiac-Implant testing datasets were visually reviewed against the underlying anatomy by a trained cardiology annotator to assess gross anatomic fidelity and failure patterns. For the Cardiac-Implant Cancer-RT testing dataset, which was not directly represented in the training datasets, lower-quartile patient-average DSC cases were evaluated using a quality-assurance review[20,21] that incorporated visual inspection and leave-one-out sensitivity analysis[22] of relative changes in mean DSC and influence ratios. Cases were determined to be “out-of-distribution” when low performance coincided with identifiable morphological or image-quality characteristics and disproportionate influence on the cohort mean DSC. Performance is reported for both the full 10-case test set and a subset removing outliers. Finally, time required for a full inference by each model is compared.

Across the full VT-SBRT cohort, clinically delivered dose distributions were retrospectively analyzed to calculate the median and range of CS mean dose ($D_{Mean}$) and maximum dose (dose to 0.03cc, $D_{0.03cc}$). For the heart and LV, doses were calculated after subtracting PTV (denoted WH-PTV and LV-PTV) to only describe dose to the uninvolved heart. Substructure-specific dose patterns were evaluated relative to PTV location using the American Heart Association 17-segment model[23] to assess target-location-dependent dose deposition. For each segment involving the target, the median substructure $D_{Mean}$ and $D_{0.03cc}$ were calculated across relevant PTVs.

# Results

## Model evaluation

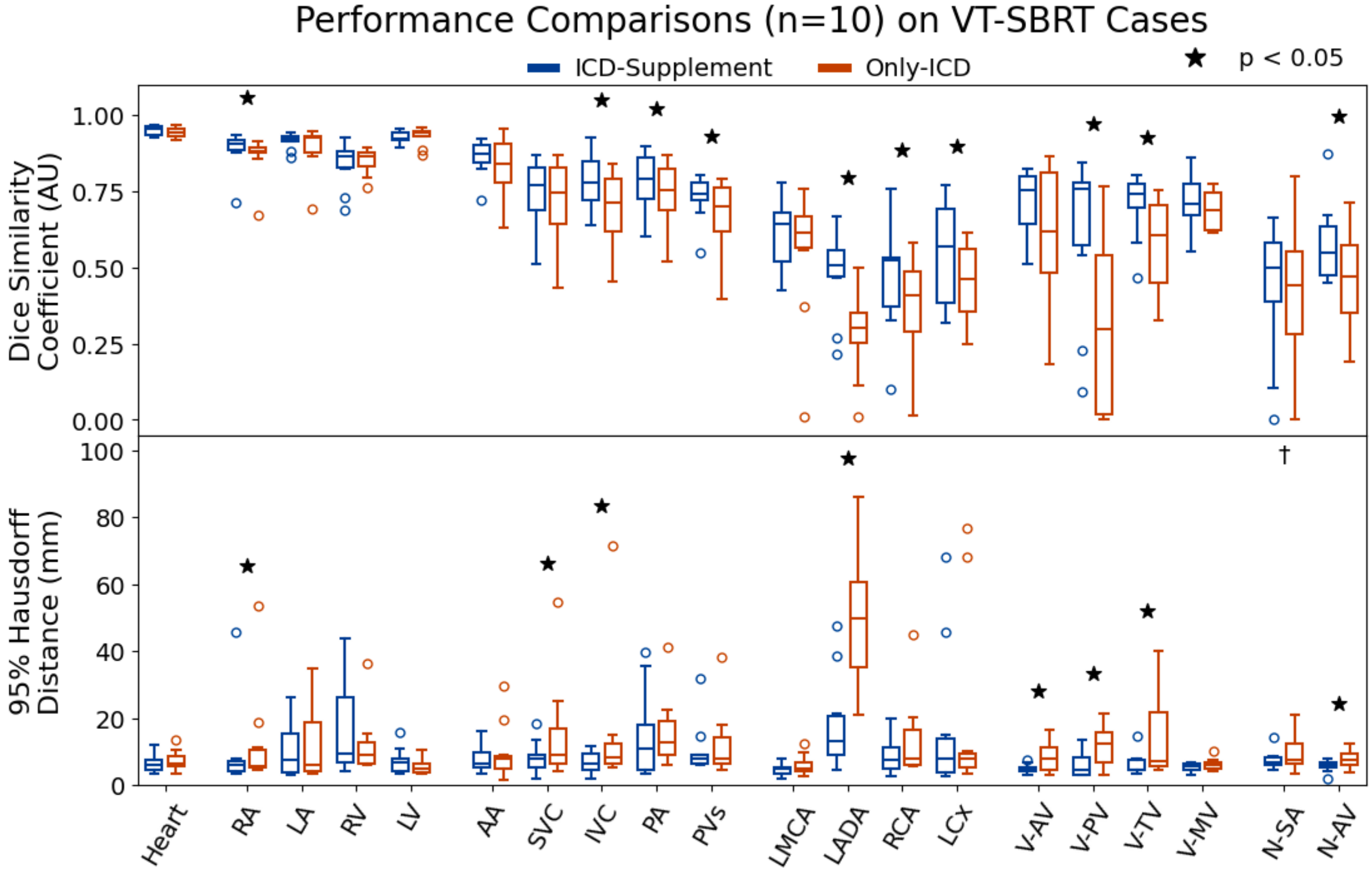

Figure 2: Quantitative model performance comparing training with implantable cardioverter-defibrillator (ICD) supplemented data against training with only ICD data for the Dice similarity coefficient and 95% Hausdorff distance (HD95) on the VT-SBRT testing dataset. Abbreviations: Arbitrary Units, AU; rest defined in text. † The model trained with ICD supplemented data failed to predict a single sinoatrial node (N-SA), resulting in infinite metrics for HD95 and was replaced with the image diagonal for comparison. On the same test case, the model trained with only ICD cases predicted with a HD95 >13cm (results not shown in graph).

The DSC and HD95 for ICD-Supplement and ICD-Only model testing on the VT-SBRT test set are shown in Figure 2. Across all substructures, ICD-Supplemented training performed all inferences in an average of 11.2±0.6 seconds with an average DSC of 0.71±0.19 (heart,

0.95±0.02; chambers, 0.89±0.06; GVs, 0.78±0.09; CAs, 0.52±0.16; valves, 0.69±0.16; nodes, 0.50±0.19) and HD95 of 10.7±22.5mm (heart, 6.6±2.7mm; chambers, 10.8±10.5mm; GVs, 9.8±7.7mm; CAs, 12.3±13.9mm; valves, 5.8±2.8mm; nodes, 21.3±64.0mm), significantly ($p<0.05$) outperforming the ICD-Only training by 0.07 increased DSC (Average DSC, 0.64±0.25) and reducing HD95 by 3.1mm (average HD95, 13.8±17.0mm). Across both metrics, adding ICD supplemental training data yielded statistically significant improvements in performance for a total of 18 individual comparisons (DSC, 10; HD95, 8). However, the model trained with ICD supplement failed to predict a single sinoatrial node and was replaced with the image diagonal, resulting in a large structure-averaged HD95. For the remaining nine cases, the average HD95 was 7.4±2.7mm. Finally, the Non-ICD model, trained on images with no ICDs, failed to generalize to this population with significantly reduced DSC of 0.54±0.28 and HD95 28.8±44.5mm ($p<0.05$, results not shown).

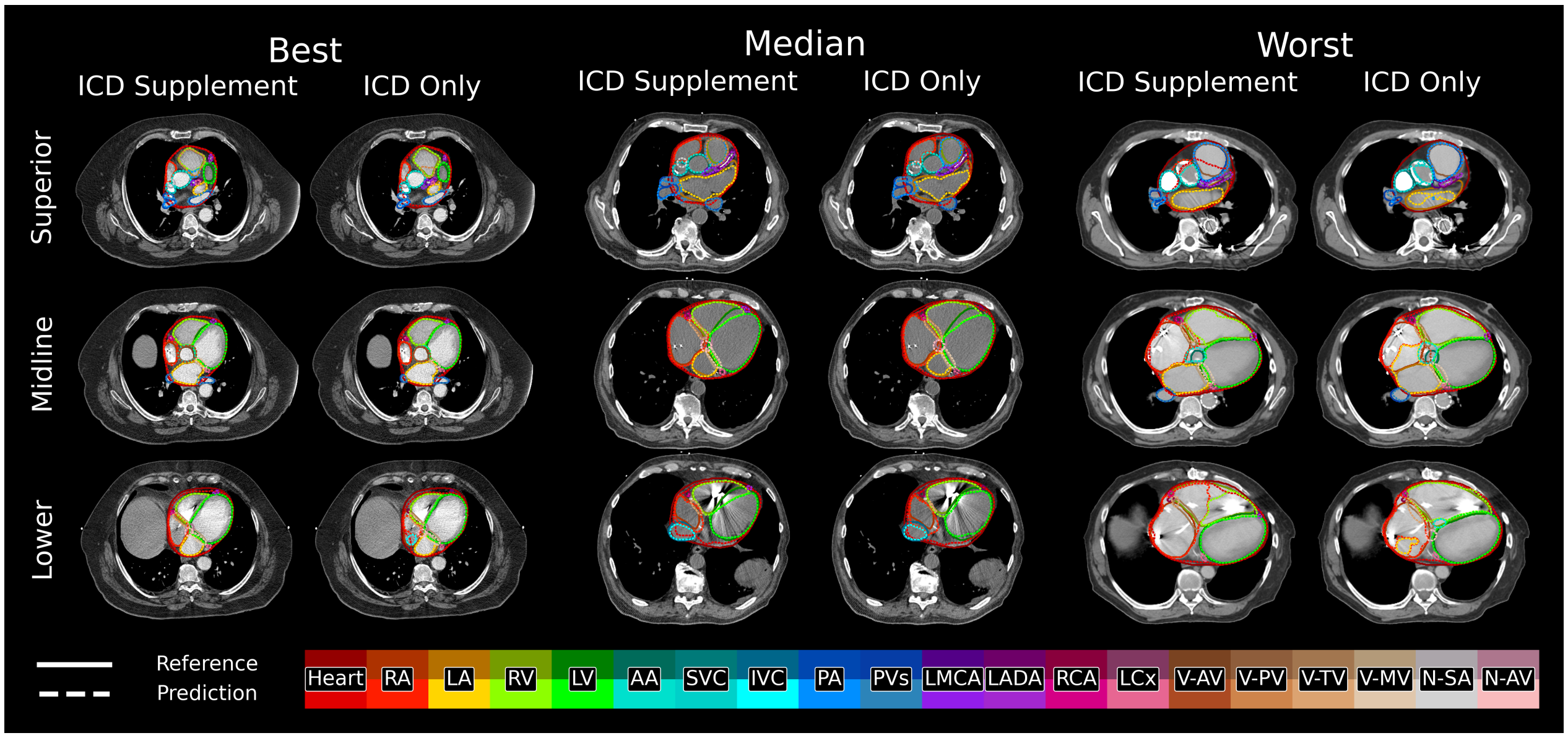

Figure 3: Qualitative analysis highlighting best, median, and worst patient-average Dice similarity coefficient across the VT-SBRT testing set, comparing the model trained with implantable cardioverter-defibrillator (ICD)-Supplemented data against the model trained with ICD-Only data. Abbreviations defined in text.

Qualitative visualization of model predictions on the VT-SBRT dataset is shown in Figure 3 for the ICD-Supplemented and ICD-Only models, highlighting best, median, and worst-case performance based on patient-average DSC by the ICD-Supplement model. In the best and median cases, the model segmented all structures with high fidelity against the reference regardless of varying contrast enhancement and presence of metal leads and artifacts. In the worst case, the model struggled with the enlarged cardiac volume (1751cc compared to ~600-700cc typical[24]) and widespread image artifacts present. Specifically, the model failed to accurately classify the right-sided structures around the right ventricle. Between methods, the model trained with ICD supplement data produced structures that more closely agreed with the underlying anatomy over the ICD only that appeared to have greater inconsistencies in boundary locations and typically did not resolve inferior CA segments.

Model evaluation of the Cardiac-Implant Cancer-RT test set performed all inferences in an average of 10.9±0.1 seconds and resulted in an average DSC of 0.68±0.23 and HD95 of 11.0±22.4mm, with three cases within the lower quartile of patient-average DSC. The lowest performance demonstrated substantial model failure including widespread false predictions and weak predictions with non-representative small CS volumes. In the leave-one-out sensitivity analysis, this case produced larger relative changes in mean DSC than the second- and third-lower quartile cases with 2.07% versus 1.55% and 0.97%, respectively, with a higher influence ratio of 1.33 versus 1.23 and 1.06 respectively. The lowest performer had a large tumor volume encompassing the left lung abutting the heart that altered the presentation of the cardiac volume, although the model still adequately accounted for the ICD leads and resolved right-sided CS. The second lowest performer demonstrated extensive high-attenuation tubular material along the CAs, potentially reflecting severe coronary calcification, coronary stenting, or other prior coronary intervention that were further affected by motion blurring and produced local image characteristics distinct from the ICD leads and substantially degraded performance for CAs. Additional prediction instability was present along the superior-inferior extents, resulting in larger disagreement between prediction location, resulting in further performance degradation. Yet, overall, the image characteristics and anatomy were largely reflective of the ICD-Supplement training set. Finally, the third lower-quartile case demonstrated structural heart disease similar to cases in the VT-SBRT dataset, including a disproportionately dilated right ventricle that increased prediction variability. However, because the second- and third-lower quartile cases demonstrated lower relative influence on cohort mean DSC and reflected anatomy represented in the intended training distribution, they were not excluded. Due to the identifiable image characteristics not represented in the training cohorts and larger influence on cohort mean DSC, the lowest performer was excluded from the final model evaluation.

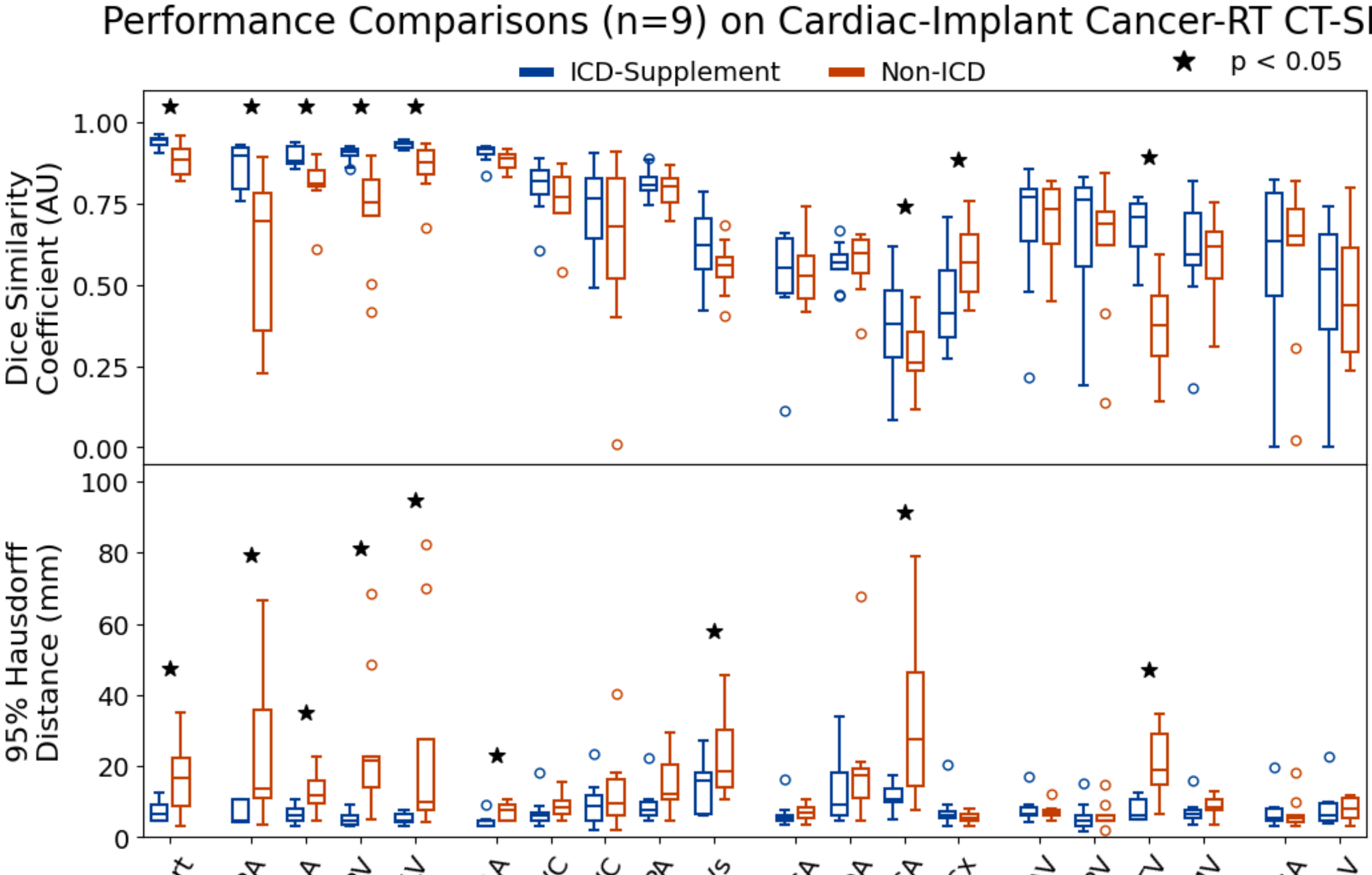

Figure 4: Quantitative model performance comparing training with implantable cardioverter-defibrillator (ICD) supplemented data against training with Non-ICD data for the Dice similarity coefficient and 95% Hausdorff distance (HD95) on the Cardiac-Implant Cancer Radiation Therapy testing dataset. Abbreviations: Arbitrary Units, AU; rest defined in text.

The DSC and HD95 for ICD-Supplement and Non-ICD model testing on the Cardiac-Implant Cancer-RT test set following review of the lower-quartile cases are shown in Figure 4. Across all substructures, ICD-Supplemented training performed with an average DSC of 0.70±0.22 (heart, 0.94±0.02; chambers, 0.90±0.04; GVs, 0.78±0.13; CAs, 0.48±0.15; valves, 0.65±0.17; nodes, 0.53±0.24) and HD95 of 7.9±5.1mm (heart, 7.5±3.0mm; chambers, 5.9±2.3mm; GVs, 9.0±6.3mm; CAs, 9.7±6.2mm; valves, 7.1±3.6mm; nodes, 7.8±5.2mm), significantly ($p<0.05$) outperforming the Non-ICD model by 0.06 increased DSC (Average DSC, 0.64±0.22) and reducing HD95 by 6.9mm (average HD95, 14.8±15.0mm). Across both metrics, adding ICD supplemental training data had statistically significant ($p<0.05$) improvements in performance for a total of 16 individual comparisons (7 cases for DSC, 9 for HD95) while the non-ICD model had one outperformance for the LCx DSC. Finally, the ICD-only model, trained on only images with ICDs from the VT-SBRT cohort, had lower performance when applied to the Cardiac-Implant Cancer-RT population with significantly reduced DSC of 0.52±0.28 and HD95 27.3±38.8mm ($p<0.05$, results not shown).

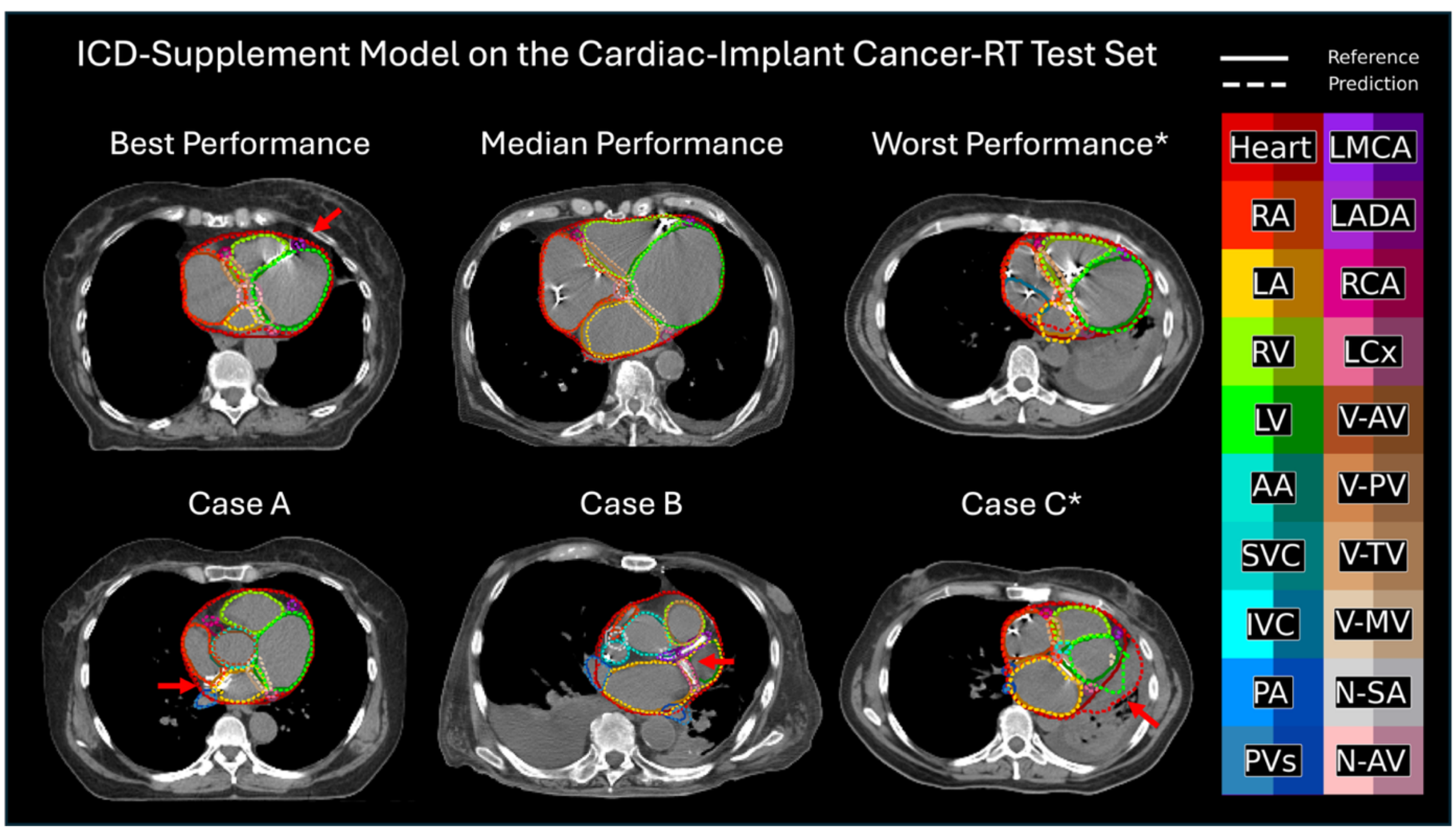

Figure 5: Qualitative performance of the model trained on implantable cardioverter-defibrillator (ICD) supplemented data demonstrating representative cases for best, median, and worst average performance (left) and three select examples highlighting performance across different patient anatomical variations (right). Case C and the worst performer, marked by an asterisk, were of the same patient and not included in the final quantitative assessment. Abbreviations defined in text.

Qualitative visualization of ICD-Supplement performance on the Cardiac-Implant, Cancer-RT test set is shown in Figure 5. Best, median, and worst model performance, based on patient-average DSC, as well as three different cases for non-ICD related characteristics are highlighted, including an example of the worst-case performer that was removed from the quantitative evaluation due to the presence of a large tumor obscuring the heart-tumor interface (Case C). In the best and median performing cases, the model is segmenting all structures and closely matching the reference contours, including in the presence of streaking artifacts and resulting signal drop out. Case A highlights a case where a patient has high-density implant related to atrial septal defect repair meanwhile Case B shows model performance in the presence of abnormal lung pathology and high-intensity values around the left CA bifurcation wherein both cases the model is unperturbed and successfully predicting local CS. In Case C, the large tumor produced widespread false positives on the patient's left side, yet the model maintained local segmentation accuracy on the right and towards the inferior regions of the heart, where ICD leads are present, while anatomy better matched the training dataset.

The DSC and HD95 for ICD-Supplement and the Non-ICD model on the non-ICD, CT-SIM dataset is shown in Figure A1. The ICD-Supplement model performed all predictions with an average inference time of 11.0±0.1 seconds and an average DSC of 0.74±0.17 and HD95 of 5.5±3.1mm. In comparison, the original, Non-ICD CT-SIM model performed with an average DSC of 0.75±0.16 and HD95 of 5.3±2.8mm as our group has previously reported[17]. While the overall DSC difference of 0.01 was statistically significant ($p<0.05$), model training with ICD-Supplemented data did not degrade performance in a clinically meaningful way. Finally, the ICD-Only model failed to generalize to non-ICD cases with a significantly reduced average DSC of 0.35±0.28 and HD95 of 75.6±81.9mm ($p<0.05$, results not shown).

## Clinical Application for VT Dosimetry

Table 1: Median and range $D_{Mean}$ and $D_{0.03cc}$ to select CS across all 19 patients. For the ventricles and heart, the planning target volume (PTV) was removed from the structure prior to calculation. Abbreviations defined in text.

| Cardiac Substructure | $D_{Mean}$ (Median, Range) Gy | $D_{0.03cc}$ (Median, Range) Gy |
|---|---|---|
| Circumflex CA | 11.2 (0.8–24.7) | 25.9 (2.1-32.3) |
| Left anterior descending CA | 8.1 (1.6-17.2) | 23.5 (3.9-32.3) |
| Ventricles - PTV | 7.7 (2.6-15.5) | 28.8 (8.0-34.4) |
| Atria | 3.9 (0.3-12.7) | 15.9 (2.0-34.8) |
| Mitral/Tricuspid Valves | 6.8 (0.6-22.9) | 15.5 (2.2-32.0) |
| Atrioventricular CN | 9.1 (0.5-27.6) | 12.6 (0.8-32.5) |
| Heart – PTV | 5.7 (2.2-8.8) | 28.8 (26.5-30.4) |

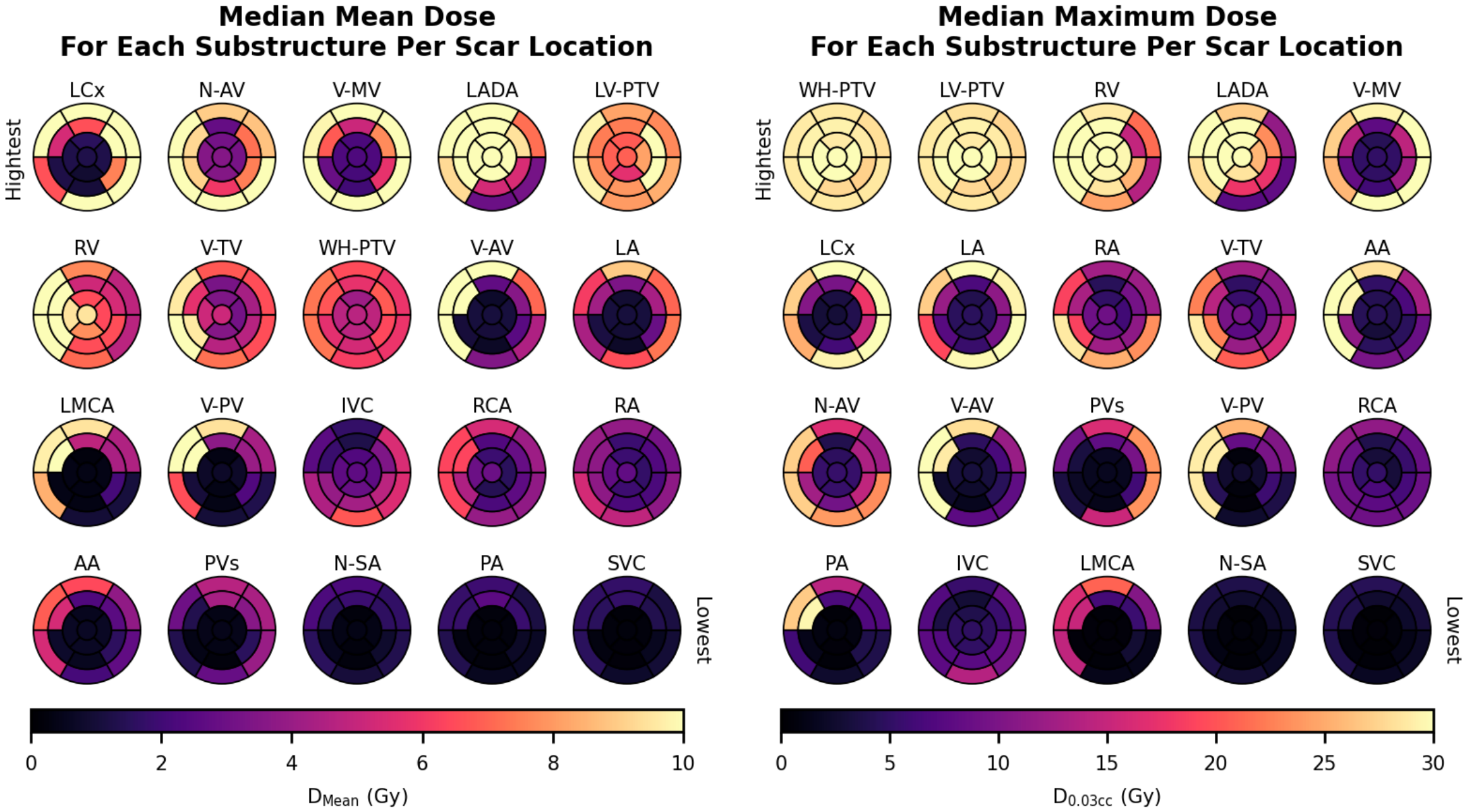

Figure 6: Scar-location-dependent substructure dose patterns summarized on the 17-segment model for mean (left) and maximum (e.g., Dose to 0.03cc, right) dose. For each segment, displayed value represents the median substructure dose across patients whose PTV involved

that segment. Structures are ordered from highest to lowest by overall median dose across the cohort. Abbreviations defined in text.

Across all 19 VT-SBRT patients, median PTV was 124.8cc (range, 50.5-306.1cc), the median number of involved LV segments (primary and secondary) was 3 (range, 1-6), and the median heart volume was 1292cc (range, 713-1997cc). CS $D_{Mean}$ and $D_{0.03cc}$ across the cohort are reported in Table 1, and scar-location-dependent dose patters are shown in Figure 6. Substructures not listed in Table 1 generally exhibited low dose levels ($D_{Mean}$ <5Gy and $D_{0.03cc}$ <8Gy), with the exception of the aorta, which had a median $D_{0.03cc}$ of 10.9 Gy. The LCx demonstrated the highest overall $D_{Mean}$, although this was localized primarily to the basal- and mid-lateral scar distributions rather than uniformly across all scar locations. In contrast, several structures, including WH-PTV, LV-PTV, RV, LADA, V-TV, and N-AV, showed persistently higher dose across a broader range of scar locations. More generally, both $D_{Mean}$ and $D_{0.03cc}$ tended to decrease toward apical scar locations for most structures. An exception was observed for several septal- or central-adjacent structures, including the PA, V-AV, V-PV, and LMCA, which demonstrated high dose primarily when the scar involved the basal- to mid-septal segments.

## Discussion

In this work, we extended a previously developed CS auto-segmentation framework[17] to enable segmentation on patients presenting to the clinic with cardiac-implants, image artifacts, and structural heart disease. By supplementing a highly curated CT-SIM dataset with a limited number of patients undergoing VT-SBRT where ICDs and structural heart disease are common, segmentation performance was significantly improved on a testing set of VT-SBRT cases (DSC, 0.71±0.19; HD95 10.7±22.5mm) while preserving performance on more standard RT CT-SIM without implants (DSC, 0.74±0.17; HD95, 5.5±3.1mm). Furthermore, the ICD-Supplemented model demonstrated generalizability to most cases in a clinically relevant cohort of cancer patients presenting with pacemakers, leadless pacemakers, ICDs, and other surgical cardiac interventions (DSC, 0.70±0.22; HD95, 7.9±5.1mm), demonstrating applicability to widespread clinical use. Finally, across the VT-SBRT cohort, involved CS dose varied widely with PTV location, indication that dose deposition is strongly target-location dependent and highlights the need for CS segmentation models that robustly delineate a broad range of CS on patients with ICDs and structural heart disease. This analysis may provide a more informative framework for studying treatment-related risk, identifying structures consistently exposed at different target locations, and guiding future efforts in selective CS sparing.

Our findings build on the recent work of van der Pol *et al.*[12], who first demonstrated the feasibility of CS segmentation on VT CT-SIM using a two-model nnU-Net ensemble for 16 structures. Van der Pol *et al.*[12] reported median DSC values of 0.89 for the chambers, 0.78 for GVs, 0.34 for CAs, and 0.35 for valves. Across the same structures, our method achieved a median DSC of 0.91 for chambers, 0.79 for GVs, 0.53 for the CAs, and 0.74 for the valves on the VT-SBRT test set. Performance differences may arise from how our work extends a previously validated, multi-modal and multi-task segmentation network that may be advantageous for cohorts where data are limited[17]. Additionally, the work presented by van der Pol *et al.*[12] utilized a multi-institutional dataset with consensus-structure definitions, whereas this work utilized contours from a single observer. Compared with prior studies performed on more conventional CT-SIM datasets, performance in our work on both the VT-SBRT and Cardiac-Implant Cancer-RT test sets were expectedly lower for some structures, likely due to the presence of ICDs and resulting artifacts, variable contrast enhancement between patients, and structural heart disease, resulting in widespread differences of structure size and location. Within that context, our results remain clinically meaningful, particularly because the model produced 20 clinically relevant CS in

~11 seconds that can be efficiently reviewed and adjusted for dose analysis on any patients with CEIDs and surgical implants present. Furthermore, the model was extended without meaningfully degrading performance on CT-SIM images without implants, preserving the efficiency and flexibility of our method to apply to a wide array of radiotherapy patients with and without implanted devices. Model training may be further improved by leveraging artifact-aware augmentation, such as introducing simulated ICD leads and metal artifact patterns into otherwise normal CT-SIM images during training[25] or by reducing artifact burden before segmentation through lead masking and image in-painting to mitigate the impact of atypical image characteristics[25].

An important limitation of this study is that, although model generalizability was improved for both VT patients and general patients undergoing RT with some form of CEIDs, challenging cases with out-of-distribution characteristics remain. When exploring generalizability to the Cardiac-Implant Cancer-RT test set, two cases were identified and removed for having lower-quartile performance associated with characteristics not included in the original training data such as. a large, proximal tumor that directly impacted the presentation of the left-sided CS (Figure 6) and the presence of severe coronary calcium, coronary stenting, or other prior coronary intervention. Furthermore, within the VT-SBRT test set, the model struggled for a patient with substantial RV dilation, leading to poor delineation and associated voxel misclassifications (Figure 3). These findings suggest that, while the model performed well across many RT patients with pacemakers, ICDs, and other surgical implants, combinations of implant-related artifacts, abnormal anatomy, and adjacent disease can still challenge segmentation robustness, particularly for the small structures like the CAs and valves. This work may be further improved by incorporating larger datasets with more diverse implants, artifacts, and anatomic presentations, and by exploring context-aware approaches, such as vision-language models, that may better contextualize and guide segmentation around the present image features[26].

A second limitation of this work is that the VT-SBRT dose assessment was a retrospective, single-institution study spanning the early development of a VT treatment program. Over that time, imaging techniques, treatment planning practice, and overall clinical experience evolved as evidence at the use of gating, contrast, and CT protocol used for treatment planning of VT. This introduced heterogeneity in image quality, treatment margins, and dose distributions across the cohort. Nevertheless, our CS model achieved comparable performance as models without ICDs[10,17] (DSC, 0.74±0.17; HD95, 5.5±3.1mm) underscoring the generalizability of our technique to different planning datasets, and offering strong potential for future clinical deployment.

Future work includes the evaluation of CS dose in patients with CEIDs or cardiac interventions undergoing thoracic radiotherapy. The ICD-Supplemented model demonstrated generalizability to the wide RT population on the Cardiac-Implant, Cancer-RT dataset (DSC, 0.70±0.22; HD95, 7.9±5.1mm). While additional data are required to further improve widespread generalizability, as previously discussed, the model successfully segmented CS in the presence of diverse CEIDs, cardiac interventions, and patient anatomies. This makes the model directly applicable to general RT dose assessment and cardiac-sparing workflows, including patients who present for standard thoracic RT with pacemakers, ICDs, or other implanted cardiac hardware. Although cardiac-sparing evaluation was outside the scope of the present work, these results support future investigations into device-adjacent CS dose assessment and opportunities for cardiac sparing in routine RT patients with cardiac implants.

## Conclusion

Implementing DL CS segmentation in the presence of CEID/ICDs for VT-SBRT and RT patients is feasible. This work demonstrated that added training data of patients with ICD present significantly improves performance when applied to general RT patients presenting with CEIDs or other cardiac intervention, further enabling CS dose assessment and cardiac sparing on patients with more complex cardiac anatomy. Furthermore, in VT-SBRT, early findings suggest dose to the CS is dependent on VT-target location and may benefit from future location-specific cardiac sparing workflows. This work extends the technical groundwork to incorporate CS in VT-SBRT and for other patients with CEIDs to support further evaluation of CS dose and development of cardiac sparing methods.

# Appendix

## Results

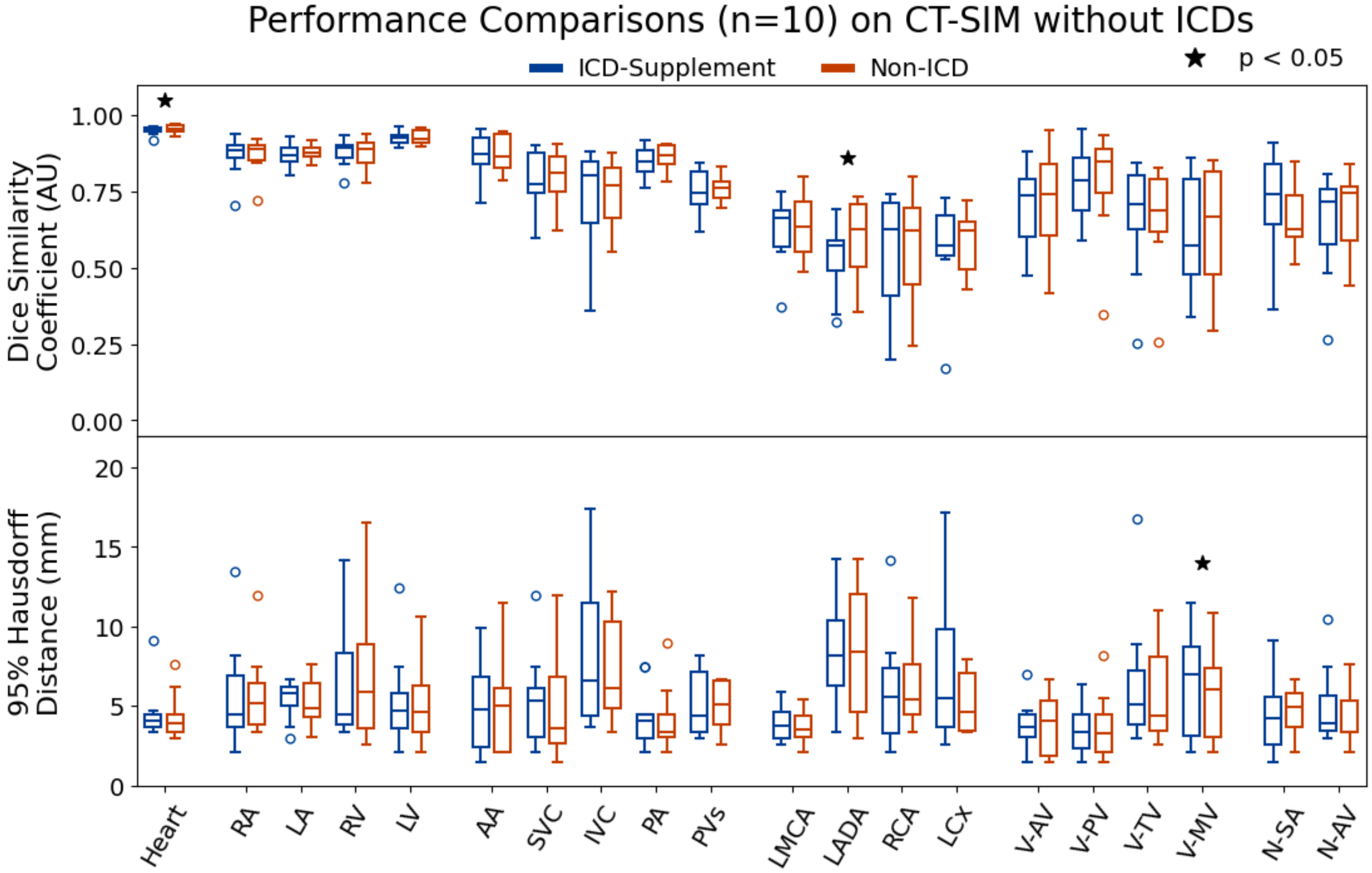

Figure A1: Quantitative model performance comparing training with implantable cardioverter-defibrillator (ICD) supplemented data against training with only cancer RT CT-SIM data for the Dice similarity coefficient and 95% Hausdorff Distance (HD95) on a testing set of CT-SIMs without ICDs present. Abbreviations: Arbitrary Units, AU; rest defined in text.